\documentclass[conference,letterpaper]{IEEEtran}
\usepackage{subfigure}
\usepackage{graphicx}
\usepackage{latexsym}
\usepackage{amsxtra} 
\usepackage{amsmath}
\usepackage{marvosym}
\usepackage{color,tabularx}
\usepackage{makecell}
\usepackage{booktabs,multicol,multirow}
\usepackage[dvipsnames]{xcolor}

\makeatletter
\def\blfootnote{\xdef\@thefnmark{}\@footnotetext}
\makeatother

\begin{document}

\title{Automated Circuit Approximation Method Driven by Data Distribution}

\author{\IEEEauthorblockN{Zdenek Vasicek, Vojtech Mrazek and Lukas Sekanina}
\IEEEauthorblockA{Brno University of Technology, Faculty of Information Technology, IT4Innovations Centre of Excellence\\
Brno, Czech Republic\\
Email: \{vasicek, imrazek, sekanina\}@fit.vutbr.cz}}


\maketitle

\begin{abstract}
We propose an application-tailored data-driven fully automated method for functional approximation of combinational circuits. We demonstrate how an application-level error metric such as the classification accuracy can be translated to a component-level error metric needed for an efficient and fast search in the space of approximate low-level components that are used in the application. This is possible by employing a weighted mean error distance (WMED) metric for steering the circuit approximation process which is conducted by means of genetic programming. WMED introduces a set of weights (calculated from the data distribution measured on a selected signal in a given application) determining the importance of each input vector for the approximation process. The method is evaluated using synthetic benchmarks and application-specific approximate MAC (multiply-and-accumulate) units that are designed to provide the best trade-offs between the classification accuracy and power consumption of two image classifiers based on neural networks.\end{abstract}

\section{Introduction}

Approximate computing exploits the fact that there are many \emph{error-resilient} applications (such as image recognition, video processing and data mining) in which \emph{quality of service} can be traded for performance or power consumption. 
Adopting the principles of approximate computing thus enables to significantly improve energy efficiency of complex computer applications.
In order to obtain an approximate implementation, a common practice is to replace selected components of the original (exact) implementation by their approximate versions. For this purpose, approximate components based on various approximation principles have been introduced (for example, see a recent survey of approximate adders and multipliers~\cite{Jiang:2017}). Even open circuit libraries nowadays provide various sorts of approximate circuits~\cite{Shafique:DAC2015,Mrazek:date17}. However, it is important to emphasize that these circuits have (almost always) been optimized with respect to general-purpose error metrics and evaluated under the assumption of uniformly distributed input values. Applying these prefabricated approximate circuits can bring some improvements in power consumption or performance, but much better trade-offs are always obtained if the approximate circuit is deliberately developed and optimized for a given application and if it exploits some knowledge about the application, for example, a particular (non-uniform) distribution of input vectors. 

These application-specific approximate circuits (ASACs) can, in principle, be obtained using automated circuit approximation methods such as ABACUS~\cite{ABACUS:DATE14}, SALSA~\cite{SALSA}, CGP~\cite{Mrazek:iccad16,Mrazek:date17} etc. if a suitable error metric is provided.
Contrasted to the manual approximation approach (represented by, e.g., ~\cite{Jiang:2017}) these methods automatically generate and evaluate candidate designs until the implementation showing acceptable trade-offs between design objectives is obtained. Let us consider an example in which the objective is to create highly efficient approximate multipliers for an image classifier based on a Convolutional Neural Network (CNN).  Papers~\cite{Mrazek:iccad16,Sarwar:date16} already shown that employing approximate multipliers optimized with respect to a given CNN can reduce (in comparison with a common truncation) the overall power consumption with a negligible impact on the accuracy. When automated approximate circuit design techniques are applied in this context, the key question is how to define the error metric for the approximation procedure working at the level of components (multipliers). It is evident that the error metrics cannot be based on the classification accuracy (i.e. at the CNN level) as obtaining this parameter requires to perform a very time consuming evaluation for each CNN containing a new candidate approximate multiplier. This approach enables to explore only a very limited number of candidate designs (within the available time) and obtain a low quality solution. On the other hand, if a common error metric 
is applied at the level of multipliers, the approximation algorithm has no way to exploit the particular data distribution observed in a given CNN.

In general, we are looking for an easy-to-calculate error metric applicable at the level of components, but providing highly correlated outputs with the quality measure used in the application containing these components. This application-tailored, but component-level error metric is then used in the circuit approximation method. 

This paper deals with an automated design of ASACs using Cartesian Genetic Programming (CGP). 
In the context of CGP-based approximations, we propose a new error metric -- a \emph{weighted mean error distance} (WMED) -- for steering the circuit approximation process. WMED introduces a set of weights (derived from the data distribution measured on a selected signal in a given application) determining the importance of each input vector for the approximation process. The principle is to allow more aggressive approximations for less important inputs (lower weights are assigned to them) and gentle approximations for highly important inputs (higher weights are assigned to them).

The proposed method is evaluated using (1) synthetic benchmark problems and (2) two instances of neural image classifiers. In the case of synthetic benchmark problems, the objective is to design an approximate multiplier $\widetilde{M}(x,y)$  showing high-quality trade-offs between WMED and power consumption.
The weights used in WMED reflect the importance of particular input vectors on $x$ input which is modeled using a probability mass function $D$. In other words, $\widetilde{M}$  is designed, optimized and approximated for a user-given $D$. This is highly relevant for applications in which one operand of the multiplier is an arbitrary input value and the importance of the second operand (roughly) follows $D$. For example, in image filters, signal filters, or artificial neurons there is always an input multiplied by a certain value (i.e. a filter coefficient or a synaptic weight) which can be statistically characterized for a given application. At the same time it is required that all multipliers have to be identical in these applications in order to obtain uniform circuit structures suitable for hardware implementation. 

In the case of neural image classifiers, application-specific approximate MAC (multiply-and-accumulate) units are designed to provide the best trade-offs between the classification accuracy and power consumption. The definition of WMED is based on the distribution of weights across all NN layers.


\section{Related work}

This paper deals with \emph{functional approximation} which is a technology-independent circuit approximation method. Its purpose is to modify the implementation (function) of a given circuit in such a way that the quality of service is kept at desired level while power consumption is reduced (or performance is increased) with respect to the original implementation. 

\subsection{Functional approximation}

Approximations have been introduced to circuits described at the transistor, gate~\cite{SALSA, Mrazek:iccad16}, register-transfer and behavioral~\cite{ABACUS:DATE14} levels. Many authors have introduced approximate operations directly at the level of abstract circuit representations such as binary decision diagrams and and-invert graphs~\cite{approx-ICCAD'16}. Basic functional approximation principles are: (i) truncation, which is based on reducing bit widths of registers and all operations of the data path; (ii) pruning, which lies in removing some parts of the circuit; (iii) component replacement, in which exact components are replaced with approximate components available in a library of approximate components; (iv) re-synthesis, in which the original logic function is replaced by a cheaper implementation; (v) other techniques such as table lookup etc. 

The automated approximation methods are often constructed as iterative methods in which many candidate approximate circuits have to be generated and evaluated.
This is, in fact, a multi-objective search process. Examples of elementary circuit modifications (i.e. steps in the search space) are replacing a gate by another one, reconnecting an internal signal or reconnecting a circuit output. It has been shown that this kind of search can effectively be performed by means of Cartesian genetic programming~\cite{miller:cgp:book,Mrazek:date17,Mrazek:iccad16}. Details on CGP will be given in Section~\ref{sec:method}.

\subsection{Approximate CNNs}

With the rapid development of artificial intelligence methods based on deep CNNs, a lot of attention has been focused on efficient hardware implementations of neural networks~\cite{sze:pieee17}. CNNs employ multiple layers of computational elements performing the convolution operation, pooling (selection/subsampling), non-linear transformations and the final classification based on a common multi-layer perceptron (MLP). 

One of the key challenges in this area is to provide fast and energy efficient \emph{inference phase} (i.e. the application of an already trained network). The reason is that trained CNNs are employed in embedded systems and have to process enormous volumes of data in a real-time scenario. As CNNs are highly error resilient, a good strategy is to reduce the bit width for all involved operations and storage elements. This approach has been taken by the Tensor Processing Unit (TPU), where only 8-bit operations are implemented in MAC units. The highly parallel processing enabled by TPU exploits a systolic array composed of 65,536 8-bit MAC units~\cite{Jouppi:2017}. 

Approximation techniques developed for circuit implementations of NNs were surveyed in~\cite{Panda:dnn16}. In the case of approximate multipliers for NNs, they are implemented either as multiplier-less multipliers~\cite{Sarwar:date16}, truncated multipliers~\cite{Jouppi:2017} or application-specific multipliers~\cite{Mrazek:iccad16}. For example, Mrazek et al. developed approximate multipliers that perform exact multiplication by zero (which is important as many weights are zero and no error is thus distributed to subsequent processing layers) and deep approximations are allowed for all the remaining operand values~\cite{Mrazek:iccad16}. On two benchmark problems, this strategy provided better trade-offs (energy vs. accuracy) than the multiplier-less multipliers~\cite{Sarwar:date16,Mrazek:iccad16}.

\section{Design of application-specific approximate circuits}
\label{sec:method}

The proposed design method based on CGP is developed for combinational circuits. For the sake of simplicity, we will focus on approximate multipliers in this section. 

\subsection{Weighted mean error distance}
We propose WMED as an extension of the conventional \emph{mean error distance} (MED).
Let $I$ and $J$ be discrete random variables representing data at the inputs of a multiplier $M$. 
Let $D$ be a probability mass function of $I$ defined as $D(x)=Pr(I=x)$.
Given $D$ and a signed approximate $w$-bit multiplier $\widetilde{M}$, WMED is defined as 
$$
\mathrm{WMED_{D}(\widetilde{M})} = {\frac{1}{2^{2w}}} \sum_{i=-2^{w-1}}^{2^{w-1} - 1} \sum_{j=-2^{w-1}}^{2^{w-1} - 1} \alpha_{i,j} \cdot | i\cdot j - \widetilde{M}(i,j)|
$$ where $\widetilde{M}(i,j)$ is the output of a signed approximate multiplier for inputs $i$  and $j$, and  $0 \leq \alpha_{i,j} \leq 1$ is the weight determined by the probability mass function $D$. In our case $\alpha_{i,j} = D(i)$ ($\sum D(i) = 1$), but a different approach can be chosen in general.
The WMED for an unsigned approximate multiplier is constructed accordingly. 
Note that $0 \leq \mathrm{WMED_{D}} \leq 1$.





\subsection{Circuit representation in CGP}

\begin{figure}[h]
\centering\includegraphics[width=.8\columnwidth]{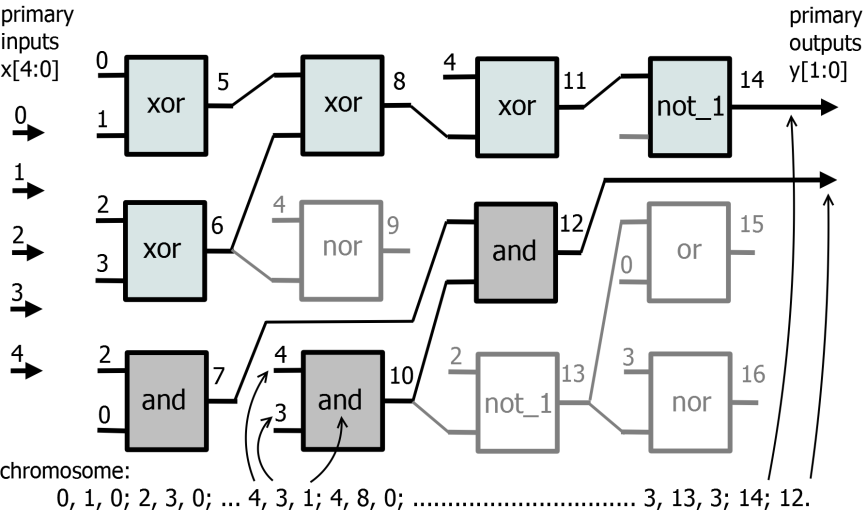}
\caption{Combinational circuit represented in CGP with parameters: $n_i$ = 5, $n_o$ = 2, $c$ = 4, $r$ = 3, $n_a$ = 2, $\Gamma$ = \{xor (encoded with 0), and (1), or (2), nor (3), not\_1 (4)\}. Nodes 9, 13, 15 and 16 are inactive.}\label{fig:cgp}
\end{figure}

In CGP~\cite{miller:cgp:book}, a combinational circuit is modeled as a two-dimensional grid of nodes (see the example in Fig.~\ref{fig:cgp}), where the type of nodes depends on the level of abstraction used in modeling (the gates are used in our case). The circuit utilizes $n_i$ primary inputs and $n_o$ primary outputs. A unique address is assigned to all primary inputs (0 -- 4 in Fig.~\ref{fig:cgp}) and to the outputs of all nodes (5 -- 16 in Fig.~\ref{fig:cgp}) to define an addressing system enabling circuit topologies to be specified. As no feedback connections are allowed in the basic version of CGP, only combinational circuits can be created. 
Each candidate circuit is represented using $S = r \times c \times (n_a + 1) + n_o$ integers, where $c$ is the number of columns, $r$ is the number of rows and $n_a$ is the maximum arity of node functions. All supported node functions are defined in the function set $\Gamma$. In this representation, the $n_a+1$ integers specify one programmable node in such a way that $n_a$ integers specify source addresses for its inputs and one integer determines the function of the node. 
This circuit representation can be seen as a netlist in which redundant components are allowed. 

\subsection{Search algorithm and fitness function}

Having a candidate circuit represented as a string of integers, new candidate circuits are created by a random modification of this string -- the so-called \emph{mutation}. 
It is important to ensure that all randomly created numbers are within a legal interval, i.e. a valid candidate circuit is always produced. 

CGP employs a simple search algorithm denoted $(1 + \lambda)$ which operates with a set of $1 + \lambda$ candidate circuits (the so-called population)~\cite{miller:cgp:book}. Starting with the original circuit (the so-called parent), a new population is created by applying the mutation operator on the original circuit and creating $\lambda$ offspring circuits. The mutation operator randomly modifies up to $h$ randomly selected integers of the string. These offspring are evaluated in terms of functionality and electrical parameters  and the so-called fitness score is assigned to them. The best performing individual is taken as a new parent. These steps are repeated until the time available for the evolution is exhausted. 

The goal of the design process is to find an approximate circuit minimizing the area on a chip and keeping WMED below a predefined threshold. The area parameter is chosen because it is highly correlated with power consumption and can quickly be estimated using the technology library (see the methodology proposed in~\cite{Mrazek:iccad16}). The design process is repeated for several target approximation errors $E_i$ in order to construct Pareto front (the error vs. the area). The fitness value $\mathbf{F}$ of a candidate approximate multiplier $\widetilde{M}$ is defined as
\begin{equation}
\mathbf{F}(\widetilde{M}) = 
\begin{cases}
A_{\widetilde{M}} & \text{if WMED}_{D}(\widetilde{M}) \le E_i  \\
\infty & \text{otherwise,} \\
\end{cases}
\label{eq:fitness}
\end{equation}
where $A_{\widetilde{M}}$ is estimated area of ${\widetilde{M}}$ and the objective is to minimize $\mathbf{F}$. 

\section{Case Study 1: Data distribution driven approximate multipliers}
\label{sec:mult_wmed}

\begin{figure}[h]
\centering\includegraphics[width=0.95\columnwidth]{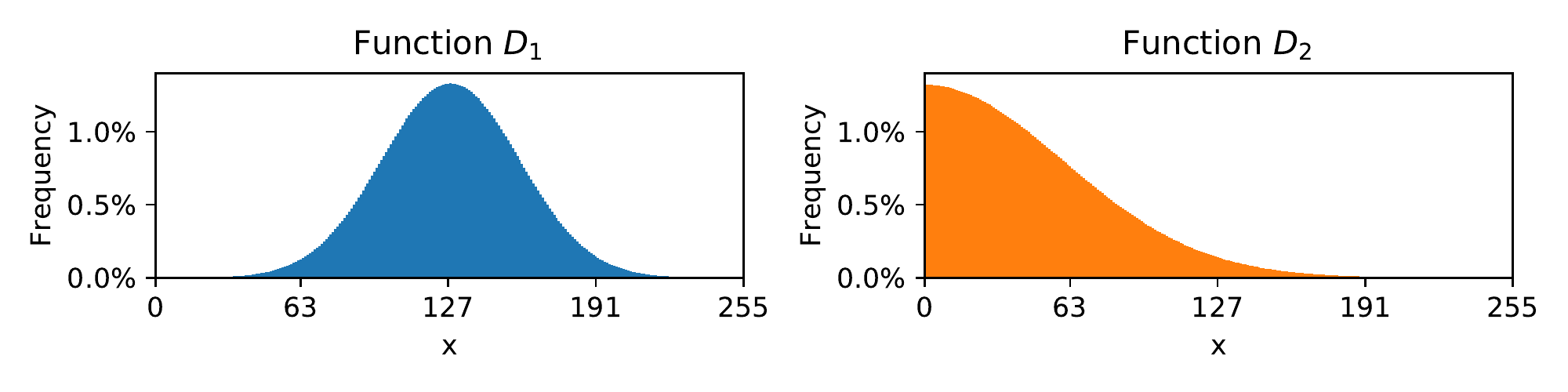}
\caption{Probability mass function $D_1$ and $D_2$}\label{fig:distributions}
\end{figure}

The objective of this section is to show that better trade-offs (between key parameters of multipliers) can be obtained in comparison with the conventional approximation methods (which are assuming uniformly distributed input data) if a non-uniform data distribution is used in the WMED definition. 
Figure~\ref{fig:distributions} shows the data distributions used in our experiments. $D_1$ and $D_2$ are arbitrarily chosen normal and half-normal distributions. The uniform distribution ($D_u$) will serve as a reference in all experiments.

\begin{figure*}[t]
\centering\includegraphics[width=0.99\textwidth]{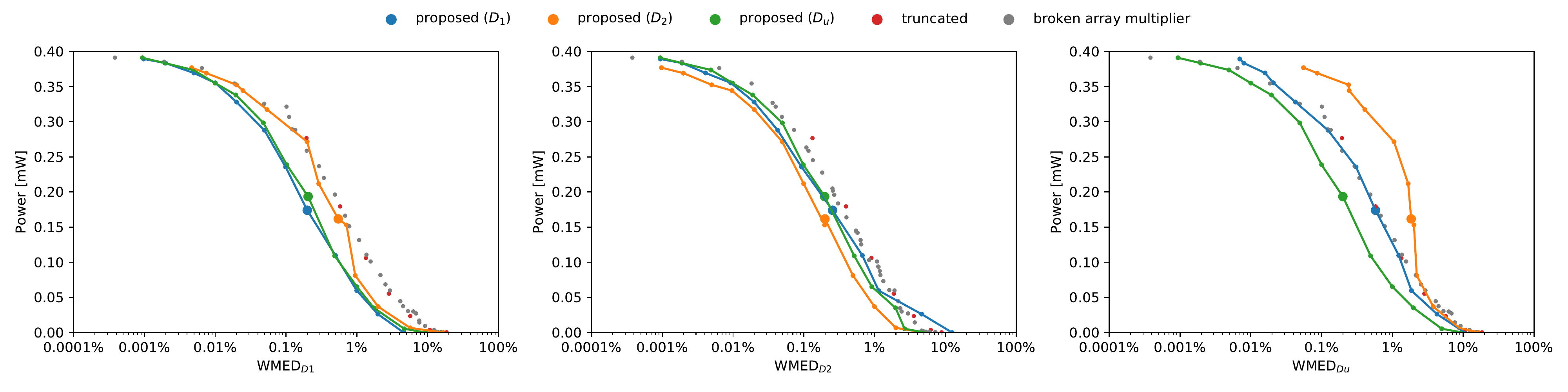}\vspace{-1em}
\caption{Parameters of approximate multipliers that were evolved according to selected distributions ($\mathrm{WMED}_{D_1}$, $\mathrm{WMED}_{D_2}$ and $\mathrm{WMED}_{D_u}$) and conventional approximate multipliers (truncated array multiplier~\cite{Jiang:2017}, broken-array multiplier~\cite{Mahdiani:TCSI2009}).}
\label{fig:mult_wmed_pareto}\vspace{-10pt}%
\end{figure*}

Approximate 8-bit multipliers are evolved using CGP which utilizes standard parameter setting as recommended in the literature~\cite{miller:cgp:book,Mrazek:iccad16}: $n_i$ = 16 (two 8-bit inputs), $n_o$ = 16, $c$ = 320 ... 490 depending on the initial multiplier, $r$ = 1, $n_a$ = 2, $\Gamma$ = \{all standard two-input gates\}, $h=5$ mutations/individual, $\lambda$ = 4.
The initial population of CGP is seeded with 
different conventional implementations of exact multipliers. The fitness function is defined according to Eq.~\ref{eq:fitness}. For all 14 target WMED values, we repeated the CGP-based design ten times (one CGP run took  1 hour). The best evolved circuits were re-synthesized with Synopsys Design Compiler (45 nm process; $V_{dd}=$1V) to obtain their power consumption and other parameters (Fig.~\ref{fig:mult_wmed_pareto}). 
In order to investigate the impact of selected distributions $D$ on properties of resulting multipliers, each multiplier is also evaluated using the remaining WMEDs that were not considered during the design.
For both $D_1$ and $D_2$ we confirmed that CGP can evolve approximate multipliers showing better trade-offs than the approximate multipliers evolved for $D_u$ and top-quality approximate multipliers available in~\cite{Jiang:2017}. 

\begin{figure}[h]\vspace{-10pt}%
\centering%
\includegraphics[width=0.99\columnwidth]{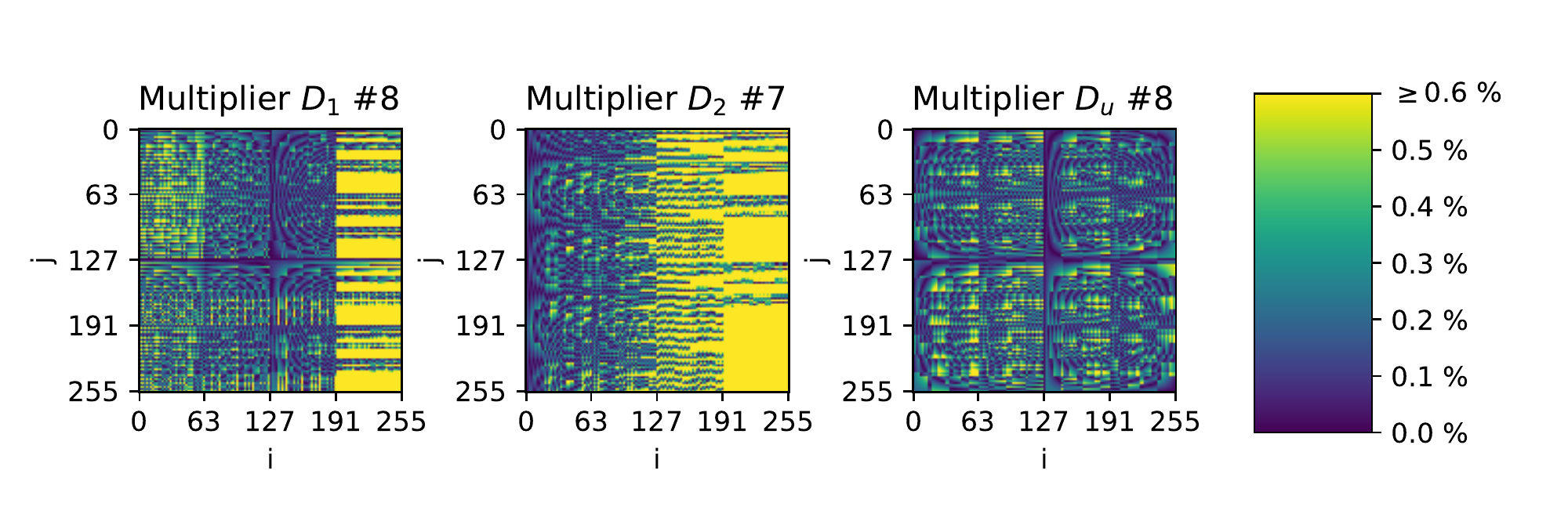}\vspace{-1em}
\caption{Approximation errors for all combinations of input vectors of selected approximate multipliers. Note that these selected multipliers are very similar in terms of power consumption and WMED.}\label{fig:heat_x}
\end{figure}

The heat maps on Fig.~\ref{fig:heat_x} show for selected multipliers (see the highlighted points in Fig.~\ref{fig:mult_wmed_pareto}) how the resulting approximation error is reflecting the data distribution applied in the approximation process. In the case of $D_1$, if the operand is around 127 the product shows a low error, but higher errors are visible for operands near to 0 and 255. In the case of $D_2$, low errors are visible for $x < 127$. In the case of $D_u$, the error is spread more uniformly.


Intuitively, approximate multipliers optimized for error distribution $D_2$ should provide better trade-offs than other multipliers when used in the image filter which is constructed to eliminate Gaussian noise. The reason is that Gaussian filters employ a $k \times k$-pixel filtering window with many close-to-zero  coefficients whose sum has to be less than 256. 
\begin{figure}[b]\vspace{-10pt}
\centering%
\includegraphics[width=0.99\columnwidth]{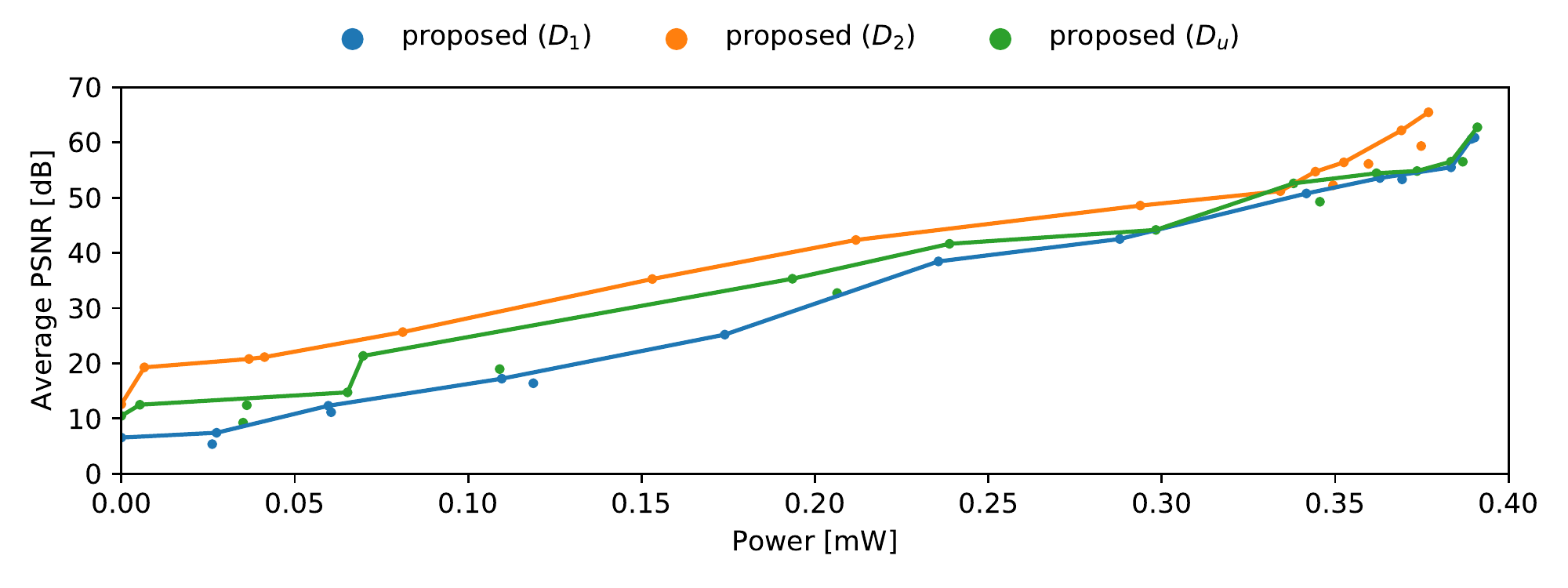}\vspace{-1em}
\caption{Average PSNR obtained using approximate Gaussian image filters employing various implementations of approximate multipliers.}\label{fig:psnr}
\end{figure}
If results of approximate multiplication by these coefficients are almost exact (the error can be arbitrarily high for non-coefficients) then the quality of filtering is higher than if the filter contains approximate multipliers showing uniformly distributed errors. Hence, we compared the impact of various approximate multipliers on the quality of filtering conducted with the approximate Gaussian filter. We used a standard Gaussian filter implementation in which $3\times3$ pixels are multiplied by nine constants.
Figure~\ref{fig:psnr} clearly shows that Gaussian filters employing approximate multipliers (which were evolved according to distribution $D_2$) show better trade-offs between Peak Signal to Noise Ratio (PSNR) and power consumption (given for the complete image filter implementation) than other implementations. PSNR is calculated as the mean value from 25 images. Please note that we have not designed any specialized approximate multipliers for this task; we just applied the approximate multipliers presented in Fig.~\ref{fig:mult_wmed_pareto}.

\section{Case Study 2: Approximate MAC units for CNNs}

When applying automated approximate circuit design techniques in the context of neural network based image classifiers, the key question is how to define an easy-to-calculate error metric for the approximation procedure working at the level of components (such as MACs and multipliers) because obtaining the classification accuracy of the whole NN is very time consuming. We will apply the CGP-based circuit approximation utilizing WMED to evolve approximate multipliers tailored for a particular trained NN. 

\subsection{Image classification benchmarks}
Our method will be evaluated in the task of image classification (digits 0 -- 9). Two NN architectures -- a popular Multi-Layer Perceptron (MLP) applied on the MNIST benchmark and CNN LeNet-5~\cite{Lecun} applied on the more challenging Google's SVHN benchmark -- will be addressed. This setup will allow us to compare our results with~\cite{Mrazek:iccad16}. 
We used the MLP network with $28 \times 28$ input neurons, 300 neurons in the hidden layer and 10 output neurons whose outputs are interpreted as the probability of each of 10 target classes. We modified LeNet-5 to be able to process $32 \times 32$ pixel images stored in SVHN. The LeNet-5 consists of five layers -- three convolution layers, two pooling layers used for data subsampling and one fully connected layer. The latter layer consists of 120 neurons outputting 10 values that are interpreted as the probability of each of 10 target classes. In LeNet-5, more than 278 thousand multiplication operations have to be executed to classify a single input image. A common MLP implementation shows 98\,\% accuracy on the MNIST data set. In the case of LeNet-5, 90.8 -- 92.7\,\% accuracy is typically reported on SVHN~\cite{Mrazek:iccad16}. 

\subsection{Reference implementation}

Common implementations of neural networks typically use a 32-bit floating-point representation of real numbers for data storage and manipulation. For both considered neural networks, we firstly apply a quantization process with Ristretto tool, which performs a fully automated trimming analysis of a given network~\cite{Ristretto}. The analysis using different bit-widths revealed that 8-bit fixed point signed values provide sufficient classification accuracy (only a 0.01\,\% resp. 0.1\,\% accuracy drop for MNIST, resp.  SVHN reported). At the end of this process, we obtained models that can be accelerated in HW using a systolic array of processing elements. Each processing element consists of an 8-bit MAC unit and $n$-bit register (such as in~\cite{Jouppi:2017}). Each MAC includes an 8-bit signed multiplier and $n$-bit adder, where $n = 8 + \mathrm{log}_2 d$ and $d$ is the maximum number of products that have to be summed up. In the case of fully connected layers and MLP, $d$ equals to the maximum number of weights that can be connected to a neuron. In the case of convolution layers, $d$ is the number of items in a kernel.

\begin{table*}[t]
\caption{Relation between WMED of best approximate multipliers and classification accuracy of approximate neural networks before and after fine-tuning. \protect\footnotemark \textup{The accuracy as well as other parameters are expressed relatively to the original NN (negative value = degradation, positive value = improvement, 0\,\% = equal to the parameters of NN when exact multipliers employed). \protect\footnotemark The design parameters are reported for the MAC units.}}
\label{tab:results}
\renewcommand\theadalign{bc}\renewcommand\theadfont{\bfseries}%
\centering
\begin{tabular}{c rrrrrrrrrr}
\toprule
\multirow{2}{*}{\thead{MAC\\WMED\\level (\%)}} &	\multicolumn{5}{c}{\bf SVHN data set} & \multicolumn{5}{c}{\bf MNIST data set} \\
\cmidrule(r){2-6}\cmidrule(l){7-11}
&	\thead{Initial accuracy} & 	\thead{After finetuning}	& \thead{PDP} & \thead{Power} &  \thead{Area}
&	\thead{Initial accuracy} & 	\thead{After finetuning}	& \thead{PDP} & \thead{Power} &  \thead{Area} \\
\midrule
0	&	0.00\,\% 	&	0.24\,\% 	&	0\,\% 	&	0\,\% 	&	0\,\% 	&	0.00\,\% 	&	0.09\,\% 	&	0\,\% 	&	0\,\% 	&	0\,\%   \\ 
0.005	&	0.02\,\%	&	0.36\,\%	&	-4\,\%	&	-8\,\%	&	-3\,\%	&	0.00\,\%	&	0.09\,\%	&	-1\,\%	&	-12\,\%	&	-3\,\%  \\ 
0.01	&	0.00\,\%	&	0.44\,\%	&	-4\,\%	&	-14\,\%	&	-5\,\%	&	0.00\,\%	&	0.10\,\%	&	-14\,\%	&	-16\,\%	&	-6\,\%  \\ 
0.05	&	0.00\,\%	&	0.51\,\%	&	-26\,\%	&	-26\,\%	&	-16\,\%	&	0.03\,\%	&	0.14\,\%	&	-28\,\%	&	-27\,\%	&	-11\,\%  \\ 
0.1	&	0.07\,\%	&	0.41\,\%	&	-29\,\%	&	-37\,\%	&	-27\,\%	&	0.05\,\%	&	0.10\,\%	&	-35\,\%	&	-32\,\%	&	-13\,\%  \\ 
0.5	&	0.08\,\%	&	0.31\,\%	&	-55\,\%	&	-57\,\%	&	-38\,\%	&	-0.01\,\%	&	0.10\,\%	&	-60\,\%	&	-65\,\%	&	-45\,\%  \\ 
1	&	0.13\,\%	&	0.20\,\%	&	-60\,\%	&	-65\,\%	&	-45\,\%	&	-0.42\,\%	&	0.12\,\%	&	-70\,\%	&	-71\,\%	&	-49\,\%  \\ 
2	&	-0.82\,\%	&	-0.41\,\%	&	-70\,\%	&	-71\,\%	&	-49\,\%	&	-4.79\,\%	&	-0.02\,\%	&	-79\,\%	&	-75\,\%	&	-53\,\%  \\ 
5	&	-18.56\,\%	&	-1.85\,\%	&	-90\,\%	&	-86\,\%	&	-70\,\%	&	-3.70\,\%	&	-0.30\,\%	&	-85\,\%	&	-83\,\%	&	-66\,\%  \\ 
10	&	-62.99\,\%	&	-5.04\,\%	&	-89\,\%	&	-87\,\%	&	-66\,\%	&	-61.14\,\%	&	-1.24\,\%	&	-91\,\%	&	-89\,\%	&	-70\,\% \\
\bottomrule
\end{tabular}
\end{table*}

\subsection{Applying available approximate circuits}

\begin{figure*}[t]
\center\includegraphics[width=0.7\textwidth]{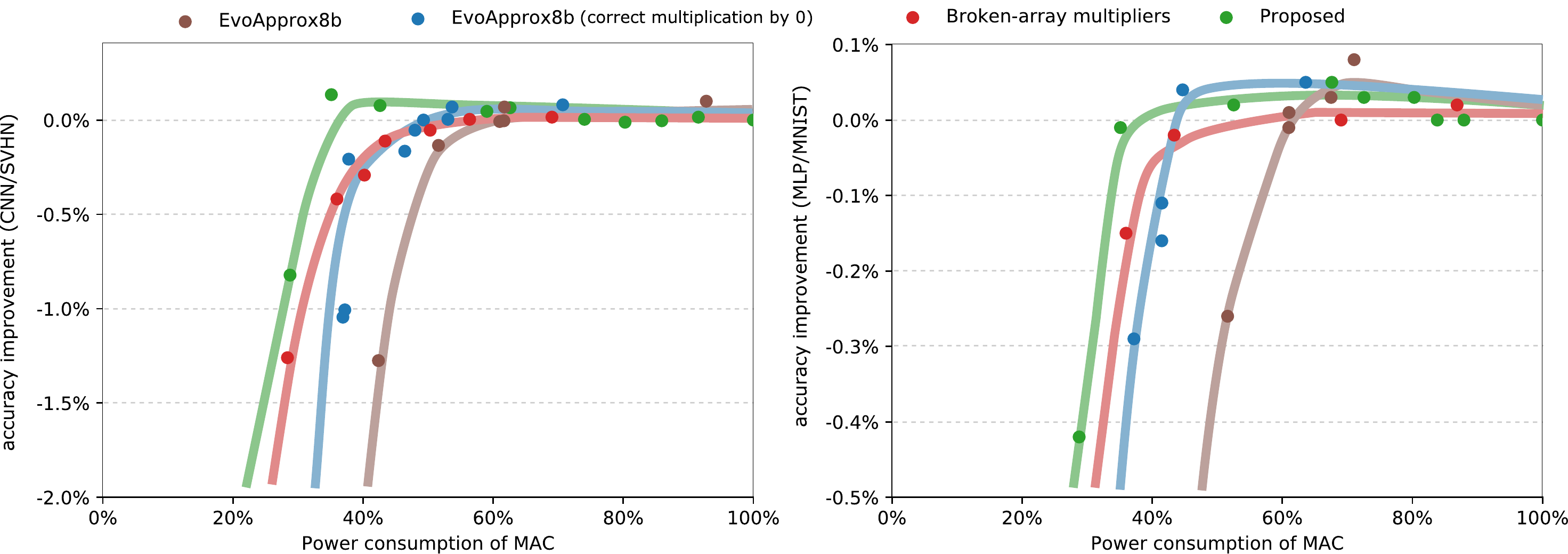}
\caption{Classification accuracy of CNN on SVHN (left) and MLP on MNIST (right) and relative power consumption when different approximate multipliers (EvoApprox8b~\cite{Mrazek:date17,Mrazek:iccad16}, broken-array multipliers~\cite{Mahdiani:TCSI2009}) are employed in MAC units. The NN accuracy is expressed relatively to the quantized model employing 8-bit accurate multiplication.
}\label{fig:comparison}
\end{figure*}

We replaced the exact multipliers with top-quality approximate 8-bit multipliers that have been proposed in literature. In particular, we considered broken-array multipliers~\cite{Mahdiani:TCSI2009} and EvoApprox8b library~\cite{Mrazek:date17}. We also utilized the approximate multipliers in which the exact multiplication by zero is guaranteed~\cite{Mrazek:iccad16}. Then we evaluated the accuracy of the neural network containing these multipliers on test data sets. Results are presented in Fig.~\ref{fig:comparison}. 

\subsection{Evolutionary design of approximate multipliers}
 
\begin{figure}[h]
\includegraphics[width=\columnwidth]{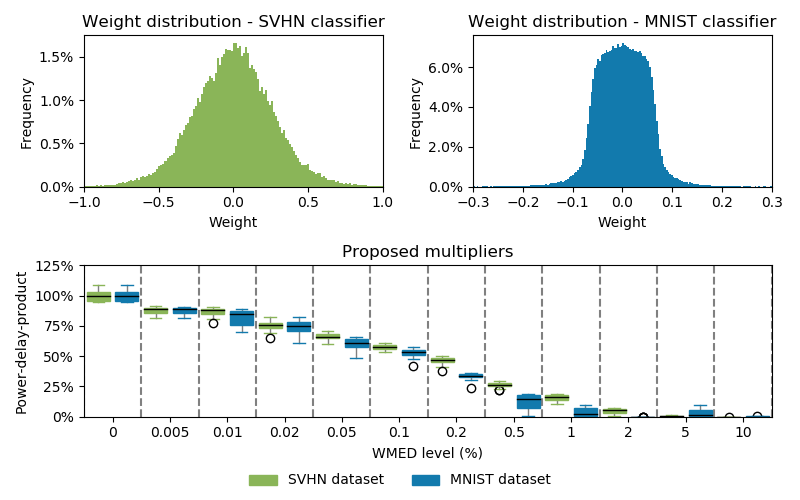}
\caption{Top: Weight distribution in neural networks trained with SVHN (left) and MNIST (right). Bottom: Relative power-delay-products of multipliers obtained from 25 independent CGP runs for a given WMED.}\label{fig:weights}
\end{figure}

We employed CGP to evolve application-tailored 8-bit approximate multipliers with the WMED error metric reflecting the properties of our target neural networks and data sets. In order to establish WMED, we analyzed the distribution of weights across all convolutional CNN layers / MLP neurons in fully trained NNs. The resulting distributions are shown in Fig.~\ref{fig:weights} (Top). In the case of SVHN, the distribution of weights is close to the normal distribution with zero mean, but MNIST has 92\,\% the most frequent values within the interval (-0.08 ... 0.08).  


CGP was used with the following parameters: $n_i$ = 16 (two 8-bit inputs), $n_o$ = 16, $c$ = 320 ... 490 depending on the initial multiplier, $r$ = 1, $n_a$ = 2, $\Gamma$ = \{all standard two-input gates\}, $h=5$ mutations/individual, $\lambda$ = 4, $10^6$ iterations/run. The fitness function is defined as proposed in Section~\ref{sec:method}.

The best discovered multipliers were integrated into MAC units and relevant design parameters were obtained with Synopsys Design Compiler (45 nm process). Fig.~\ref{fig:weights} (Bottom) shows Power Delay Product (PDP) by means of box plot graphs for resulting approximate multipliers evolved for desired WMED. Each box plot was constructed from 25 independent CGP runs. For example, if WMED is constrained to 0.2\%, PDP can be reduced by 50\,\% in the case of LeNet-5 on SVHN.

\subsection{Integration of approximate MACs to CNNs}

The best non-dominated MACs were integrated to both neural network architectures whose classification accuracy was then calculated using test sets (Table 1). We can observe that CNN accuracy remains practically unchanged for WMED~$\leq$~0.5\%. However, corresponding PDP of MAC units was reduced by 55\,\%. If a deeper approximation is allowed (WMED = 2\%), a 70\,\% reduction of PDP is reported. 

The fine-tuning of the NN weights can, in principle, improve the accuracy drop introduced by quantization. During this fine-tuning, the network learns how to classify images with approximate multipliers. Table~\ref{tab:results} shows that the effect of fine-tuning (10 iterations employed) is enormous especially in the case of 5\,\% and 10\,\% error. For 10\,\% error, for example, the accuracy was improved from $-62.99$\,\% to $-5.04$\,\% for SVHN and from $-61.14$\,\% to $-1.24$\,\% for MNIST. As it is acceptable to tolerate a 1\,\% accuracy drop in practice, we can achieve more than 70\,\% power and PDP reduction for SVHN (WMED = 2\%) and 85\,\% reduction for MNIST (WMED = 5\%). Fig.~\ref{fig:comparison} compares the classification accuracy (obtained by LeNet-5 on SVHN and MLP on MNIST) and relative power consumption when different approximate multipliers are employed in MAC units. Solutions obtained with the proposed method are clearly dominating. 

\section{Conclusions}

By means of the proposed error metric -- WMED -- we demonstrated how an application-level error metric can be translated to a component level and exploited in searching for high quality application-specific approximate circuits. The method has been evaluated in the design of approximate multipliers in which the importance of one of the operands is determined using a probability mass function. Under this scenario we evolved approximate multipliers showing better trade-offs than (i) approximate multipliers evolved with a common error metric and (ii) high-quality conventionally designed approximate multipliers. The impact of the method was demonstrated in the approximate implementation of Gaussian image filters. We also showed that when evolved MAC units are used in NN-based classifiers, 65\,\% power consumption reduction is obtained (in the MAC units), with a negligible impact on the accuracy of classification.  

\vspace{3pt}
\emph{This work was supported by Czech Science Foundation project 19-10137S.}

\bibliographystyle{IEEEtran}
\bibliography{IEEEabrv,date19w}

\begin{thebibliography}{10}
\providecommand{\url}[1]{#1}
\csname url@samestyle\endcsname
\providecommand{\newblock}{\relax}
\providecommand{\bibinfo}[2]{#2}
\providecommand{\BIBentrySTDinterwordspacing}{\spaceskip=0pt\relax}
\providecommand{\BIBentryALTinterwordstretchfactor}{4}
\providecommand{\BIBentryALTinterwordspacing}{\spaceskip=\fontdimen2\font plus
\BIBentryALTinterwordstretchfactor\fontdimen3\font minus
  \fontdimen4\font\relax}
\providecommand{\BIBforeignlanguage}[2]{{%
\expandafter\ifx\csname l@#1\endcsname\relax
\typeout{** WARNING: IEEEtran.bst: No hyphenation pattern has been}%
\typeout{** loaded for the language `#1'. Using the pattern for}%
\typeout{** the default language instead.}%
\else
\language=\csname l@#1\endcsname
\fi
#2}}
\providecommand{\BIBdecl}{\relax}
\BIBdecl

\bibitem{Jiang:2017}
H.~Jiang, C.~Liu \emph{et~al.}, ``A review, classification, and comparative
  evaluation of approximate arithmetic circuits,'' \emph{J. Emerg. Technol.
  Comput. Syst.}, vol.~13, no.~4, Aug. 2017.

\bibitem{Shafique:DAC2015}
M.~Shafique, W.~Ahmad, R.~Hafiz, and J.~Henkel, ``A low latency generic
  accuracy configurable adder,'' in \emph{Proceedings of the 52nd Annual Design
  Automation Conference}.\hskip 1em plus 0.5em minus 0.4em\relax ACM, 2015, pp.
  86:1--86:6.

\bibitem{Mrazek:date17}
V.~Mrazek, R.~Hrbacek, Z.~Vasicek, and L.~Sekanina, ``Evoapprox8b: Library of
  approximate adders and multipliers for circuit design and benchmarking of
  approximation methods,'' in \emph{Design, Automation {\&} Test in Europe
  Conference {\&} Exhibition, {DATE} 2017}, 2017, pp. 258--261.

\bibitem{ABACUS:DATE14}
K.~Nepal, Y.~Li, R.~I. Bahar, and S.~Reda, ``{ABACUS}: A technique for
  automated behavioral synthesis of approximate computing circuits,'' in
  \emph{Proceedings of the Conference on Design, Automation and Test in
  Europe}, ser. DATE'14.\hskip 1em plus 0.5em minus 0.4em\relax EDA Consortium,
  2014, pp. 1--6.

\bibitem{SALSA}
S.~Venkataramani, A.~Sabne, V.~J. Kozhikkottu, K.~Roy, and A.~Raghunathan,
  ``{SALSA}: systematic logic synthesis of approximate circuits,'' in \emph{The
  49th Design Automation Conference}.\hskip 1em plus 0.5em minus 0.4em\relax
  ACM, 2012, pp. 796--801.

\bibitem{Mrazek:iccad16}
V.~Mrazek, S.~S. Sarwar, L.~Sekanina, Z.~Vasicek, and K.~Roy, ``Design of
  power-efficient approximate multipliers for approximate artificial neural
  networks,'' in \emph{Proceedings of the IEEE/ACM International Conference on
  Computer-Aided Design}.\hskip 1em plus 0.5em minus 0.4em\relax ACM, 2016, pp.
  811--817.

\bibitem{Sarwar:date16}
S.~S. Sarwar, S.~Venkataramani, A.~Raghunathan, and K.~Roy, ``Multiplier-less
  artificial neurons exploiting error resiliency for energy-efficient neural
  computing,'' in \emph{Proc. of the Design, Automation \& Test in Europe
  Conference}.\hskip 1em plus 0.5em minus 0.4em\relax EDA Consortium, 2016, pp.
  1--6.

\bibitem{approx-ICCAD'16}
A.~Chandrasekharan, M.~Soeken, D.~Gro{\ss}e, and R.~Drechsler,
  ``Approximation-aware rewriting of aigs for error tolerant applications,'' in
  \emph{Proc. of ICCAD'16}.\hskip 1em plus 0.5em minus 0.4em\relax ACM, 2016,
  pp. 83:1--83:8.

\bibitem{miller:cgp:book}
J.~F. Miller, \emph{Cartesian Genetic Programming}.\hskip 1em plus 0.5em minus
  0.4em\relax Springer-Verlag, 2011.

\bibitem{sze:pieee17}
V.~Sze, Y.~Chen, T.~Yang, and J.~S. Emer, ``Efficient processing of deep neural
  networks: A tutorial and survey,'' \emph{Proceedings of the IEEE}, vol. 105,
  no.~12, pp. 2295--2329, 2017.

\bibitem{Jouppi:2017}
N.~P. Jouppi, C.~Young, N.~Patil \emph{et~al.}, ``In-datacenter performance
  analysis of a tensor processing unit,'' in \emph{Proc. of the 44th Annual
  Int. Symposium on Computer Architecture}.\hskip 1em plus 0.5em minus
  0.4em\relax ACM, 2017, pp. 1--12.

\bibitem{Panda:dnn16}
P.~Panda, A.~Sengupta, S.~S. Sarwar, G.~Srinivasan, S.~Venkataramani,
  A.~Raghunathan, and K.~Roy, ``Invited -- cross-layer approximations for
  neuromorphic computing: From devices to circuits and systems,'' in \emph{53nd
  Design Automation Conference}.\hskip 1em plus 0.5em minus 0.4em\relax IEEE,
  2016, pp. 1--6.

\bibitem{Mahdiani:TCSI2009}
H.~R. Mahdiani, A.~Ahmadi, S.~M. Fakhraie, and C.~Lucas, ``Bio-inspired
  imprecise computational blocks for efficient vlsi implementation of
  soft-computing applications,'' \emph{IEEE Transactions on Circuits and
  Systems I: Regular Papers}, vol.~57, no.~4, pp. 850--862, April 2010.

\bibitem{Lecun}
Y.~{LeCun}, L.~Bottou, Y.~Bengio, and P.~Haffner, ``Gradient-based learning
  applied to document recognition,'' \emph{Proceedings of the IEEE}, vol.~86,
  no.~11, pp. 2278--2324, 1998.

\bibitem{Ristretto}
P.~Gysel, J.~Pimentel, M.~Motamedi, and S.~Ghiasi, ``Ristretto: A framework for
  empirical study of resource-efficient inference in convolutional neural
  networks,'' \emph{IEEE Transactions on Neural Networks and Learning Systems},
  vol.~29, no.~11, pp. 5784--5789, 2018.

\end{thebibliography}

\end{document}